# Ciphertext policy Attribute based Encryption with anonymous access policy

A.Balu[1], K.Kuppusamy[2]

[1] Research Associate, [2] Associate Professor
Department of Computer Science & Engg.,Alagappa University, Karaikudi,
Tamil Nadu, India.
balusuriya@yahoo.co.in, kkdiksamy@yahoo.com

*Abstract*

*In Ciphertext Policy Attribute based Encryption scheme, the encryptor can fix the policy, who can decrypt the encrypted message. The policy can be formed with the help of attributes. In CP-ABE, access policy is sent along with the ciphertext. We propose a method in which the access policy need not be sent along with the ciphertext, by which we are able to preserve the privacy of the encryptor. The proposed construction is provably secure under Decision Bilinear Diffe-Hellman assumption.*

## 1   INTRODUCTION

Recently, much attention has been attracted by a new public key primitive called Attribute-based encryption (ABE). ABE has significant advantage over the traditional PKC primitives as it achieves flexible one-to-many encryption instead of one-to-one. ABE is envisioned as an important tool for addressing the problem of secure and fine-grained data sharing and access control. In an ABE system, a user is identified by a set of attributes. In their seminal paper Sahai and Waters [6] use biometric measurements as attributes in the following way. A secret key based on a set of attributes $\omega$, can decrypt a ciphertext encrypted with a public key based on a set of attributes $\omega'$, only if the sets $\omega$ and $\omega'$ overlap sufficiently as determined by a threshold value t. A party could encrypt a document to all users who have certain set of attributes drawn from a pre-defined attribute universe. For example, one can encrypt a blood group wanted document to all donors of that specific blood group from a particular locality of specific age group. In this case the document would be encrypted to the attribute subset {"B$^+$", "Karaikudi", "Age 20-25"}, and only users with all of these three attributes in the blood bank can hold the corresponding private keys and thus decrypt the document, while others cannot.

There are two variants of ABE: Key-Policy based ABE (KP-ABE) [5] and Ciphertext Policy based ABE(CP-ABE) [1,2,3,4]. In KP-ABE, the ciphertext is associated with a set of attributes and the secret key is associated with the access policy. The encryptor defines the set of descriptive attributes necessary to decrypt the ciphertext. The trusted authority who generates user's secret key defines the combination of attributes for which the secret key can be used. In CP-ABE, the idea is reversed: now the ciphertext is associated with the access policy and the encrypting party determines the policy under which the data can be decrypted, while the secret key is associated with a set of attributes.

Besides fine-grained access policy, there is an increasing need to protect user privacy in today's access control systems. In some critical circumstances, the access policy itself could be sensitive information. Therefore, we propose an attribute – based encryption scheme where encryptor specified access policies are hidden. Even the legitimate decryptor cannot obtain the information about the access policy associated with the encrypted data more than the fact that she can decrypt the data.

### Our Contribution

We present a scheme for constructing a Ciphertext Policy Attribute based Encryption with hidden access policy and provide security under the Decisional Diffie-Hellman assumption. In our scheme access policy can be expressed using AND, OR boolean operators, so that it is possible to express the access policy effectively. Each attribute $a_i$ in the access policy can take multiple values. The access policy can be represented by an n-ary tree, the leaf nodes represents the attribute present in the access policy, interior nodes represents the AND, OR operators. Each attribute in the leaf node can take multiple values. The value assigned for the leaf node by the secret sharing method will be distributed to these multiple values. In our scheme, it is not necessary to put all the attributes in the access policy.

### Related Work

Since the introduction of ABE in implementing fine-grained access control systems, a lot of works have been proposed to design flexible ABE schemes. There are two methods to realize the fine-grained access control based on ABE: KP-ABE and CP-ABE. They were both mentioned in [4] by Goyal et al. In KP-ABE, each attribute private key is associated with an access structure that specifies which type of ciphertexts the key is able to decrypt, and ciphertext is labeled with sets of attributes. In a CP-ABE system, a user's key is associated with a set of attributes and an ecrypted ciphertext will specify an access policy over attributes. The first KP-ABE construction [4] realized the monotonic access structures for key policies. Bethencourt et al. [2] proposed the first CP-ABE construction. The construction [2] is only proved secure under the generic group model. To overcome this weakness, Cheung and Newport [3] presented another construction that is proved to be secure under the standard model. To achieve receiver-anonymity, Boneh and Waters [10] proposed a predicate encryption scheme based on the primitive called Hidden Vector Encryption.

The first Anonymous Ciphertext policy Attribute-Based Encryption (ABE) construction was introduced by Nishide et al. [7]. They gave two CP-ABE schemes with partially hidden ciphertext policies in the sense that possible values of each attribute in the system should be known to an encryptor in advance and the encryptor can hide what subset of possible values for each attribute in the ciphertext policy can be used for successful decryption. The policy can be expressed as AND gates on multi valued attributes with wild cards. They describe their constructions in the multi-valued attribute setting where an attribute can take multiple values. The legitimate decryptor cannot obtain the information about the ciphertext policy. The Second construction was proposed by Keita Emura et al. [8] focusing Key anonymity with respect to the authority. In this model, even if an adversary has the master key, the adversary cannot guess what identity is associated with the ciphertext. The access structure used in their scheme is restricted to an AND gate only. Third construction was proposed by Jin li et al.[9] gave accountable, anonymous Ciphertext policy attribute based encryption. This is achieved by binding users identity in the attribute private key. They gave two constructions, one with short public parameters and the other with short ciphertext. They use two different generators to prevent the public verifiability of the ciphertext validity, which achieves hidden policy. In this method also the access structure can be specified as AND gate of multi valued attributes.

## 2 PRELIMINARIES

### 2.1 Bilinear Maps

Let $G$ and $G_1$ be two multiplicative cyclic groups of prime order p. Let g be a generator of $G$ and e be a bilinear map, $e : G \times G \rightarrow G_1$. The bilinear map e has the following properties:

1. Bilinearity : for all $u,v \in G$ and $a,b \in Z_p$, we have
   $e(u^a, v^b) = e(u,v)^{ab}$
2. Non-degeneracy : $e(g,g) \neq 1$.

We say that G is a bilinear group if the group operation in G and bilinear map $e : G \times G \rightarrow G_1$ are both efficiently computable. Notice that the map e is symmetric since $e(g^a, g^b) = e(g,g)^{ab} = e(g^b, g^a)$.

### Decisional Bilinear Diffie-Hellman Assumption

A challenger chooses a group G of prime order p according to the security parameter. Let $a,b,c \in Z_p$ be chosen at random and g be a generator G. The adversary when given $(g, g^a, g^b, g^c)$ must distinguish a valid tuple $e(g,g)^{abc} \in G_1$ from a random element R in $G_1$.

An algorithm $\mathcal{A}$ that outputs $\{0,1\}$ has advantage $\epsilon$ in solving decisional BDH if
$|\Pr[\mathcal{A}(g, g^a, g^b, g^c, D=e(g,g)^{abc}) = 0] - \Pr[\mathcal{A}(g, g^a, g^b, g^c, D=R) = 0]| \geq \epsilon$

**Definition 1** The DBDH assumption holds if no polytime algorithm has a non-negligible advantage in solving the DBDH problem.

## 2.2 Access structure

**Definition 2** Let $U = \{a_1, a_2, ..., a_n\}$ be a set of attributes. For $a_i \in U$, $S_i = \{v_{i,1}, v_{i,2}, ..., v_{i,n_i}\}$ is a set of possible values, where $n_i$ is the number of possible values for $a_i$. Let $L = [L_1, L_2, ..., L_n]$ $L_i \in S_i$ be an attribute list for a user, and $W = [W_1, W_2, ..., W_n]$ $W_i \in S_i$ be an access policy. The notation $L \models W$ express that an attribute list L satisfies an access policy W, namely $L_i = W_i$ (i=1,2..,n). The notation $L \not\models W$ implies L not satisfying the access structure W.

## 2.3 Ciphertext Policy Attribute based Encryption

A cipher text policy attribute based encryption scheme consists of four fundamental algorithms: Setup, Key Generation, Encryption and Decryption.

**Setup**: The setup algorithm takes no input other than the implicit security parameter. It outputs the public parameters PK and a master key MK.

**Key Generation** (MK,S): The key generation algorithm takes as input the master key MK and a set of attributes S that describe the key. It outputs a private key SK.

**Encrypt** (PK,A, M): The encryption algorithm takes as input the public parameters PK, a message M, and an access structure A over the universe of attributes. The algorithm will encrypt M and produce a ciphertext CT such that only a user that possesses a set of attributes that satisfies the access structure will be able to decrypt the message. Assume that the ciphertext implicitly contains A.

**Decrypt**(PK,CT,SK): The decryption algorithm takes as input the public parameters PK, a ciphertext CT, which contains an access policy A, and a private key SK, which is a private key for a set S of attributes. If the set S of attributes satisfies the access structure A then the algorithm will decrypt the ciphertext and return a message M.

## 2.4 Security Model for CP-ABE

**Init.** The adversary sends the two different challenge access structures $W_0^*$ and $W_1^*$ to the challenger.

**Setup**. The challenger runs the Setup algorithm and gives the public parameters, PK to the adversary.

**Phase 1**. The adversary sends an attribute list L to the challenger for a Key Gen query, where ($L \not\models W_0^*$ and $L \not\models W_1^*$) or ($L \models W_0^*$ and $L \models W_1^*$) The challenger answers with a secret key for these attributes.

**Challenge.** The adversary submits two equal length messages $M_0$ and $M_1$. Note that if the adversary has obtained $SK_L$ where ($L \models W_0^*$ and $L \models W_1^*$) then $M_0 = M_1$. The challenger chooses d randomly from {0,1} and runs Encrypt(PK, $M_d$, $W_d^*$). The challenger gives the ciphertext CT* to the adversary.

**Phase 2**. Same as Phase 1.

**Guess.** The adversary outputs a guess d' of d.

The advantage of an adversary A in this game is defined as $\Pr[d'=d] - \frac{1}{2}$.

**Definition 3** A ciphertext-policy attribute based encryption scheme is secure if all polynomial time adversaries have at most a negligible advantage in the above game.

## 3 CONSTRUCTION

Proposed solution consists of 4 phases, Setup Phase, Key Generation Phase, Encryption Phase and Decryption Phase.

**Set Up:**

The setup algorithm chooses a group G of prime order p and a generator g.

Step 1: A trusted authority generates a tuple $G=[p,G,G_1,g \in G, e] \leftarrow Gen(1^k)$.

Step 2: For each attribute $a_i$ where $1 \leq i \leq n$, the authority generates random value $\{a_{i,t} \in Z_p^*\}$ $1 \leq t \leq n_i$ and computes $\{T_{i,t} = g^{a_{i,t}}\}$ $1 \leq t \leq n_i$

Step 3: Compute $Y = e(g,g)^\alpha$ where $\alpha \in Z_p^*$

Step 4: The public key PK consists of $[Y,p,G,G_1,e,\{\{T_{i,t}\} 1 \leq t \leq n_i\} 1 \leq i \leq n]$

The master key Mk is $[\alpha, \{\{a_{i,t} \in Z_p^*\} 1 \leq t \leq n_i\} 1 \leq i \leq n]$

**Key Generation (MK,L):** The Key Generation algorithm takes master key MK and the attribute list of the user as input and do the following

Let $L=[L_1,L_2,\ldots,L_n]=\{v_{1,t_1}, v_{2,t_2}, \ldots, v_{n,t_n}\}$ be the attribute list for the user who obtain the corresponding secret key.

Step1: The trusted authority picks up random values $\lambda_i \in Z_p^*$ for $1 \leq i \leq n$ & $r \in Z_p^*$ and computes $D_0 = g^{\alpha - r}$.

Step2: For $1 \leq i \leq n$ the authority also computes $D_{i,1}, D_{i,2} = [g^{r+\lambda_i a_{i,t}}, g^{\lambda_i}]$ where $L_i = v_{i,t_i}$ The secret key is $[D_0, D_{i,1}, D_{i,2}]$.

**Encrypt(PK,M,W):** An encryptor encrypts a message $M \in G_1$ under a cipher text policy $W=[w_1,w_2,..,w_n]$ and proceed as follows.

Step1 : Select $s \in Z_p^*$ and compute $C_0=g^s$ and $C^\sim = M \cdot Y^s = M \cdot e(g,g)^{\alpha s}$

Step2: Set the root node of W to be s, mark all child nodes as un-assigned, and mark the root node assigned.

Recursively, for each un-assigned non leaf node, do the following

a) If the symbol is ∧ and its child nodes are unassigned, we assign a random value $s_i$, $1 \leq s_i \leq p-1$ and to the last child node assign the value

$s_t = s - \sum_{i=1}^{t-1} s_i \bmod p$. Mark this node assigned.

b) If the symbol is ∨, set the values of each node to be s. Mark this node assigned.
c) Each leaf attribute $a_i$, can take any possible multi values, the value of the share $s_i$ is distributed to those values and compute

   $[C_{i,t,1}, C_{i,t,2}] = [g^{s_i}, T_{i,t}^{s_i}]$. The cipher text CT is
   $[C\tilde{~}, C_0, \{\{C_{i,t,1}, C_{i,t,2}\} 1 \leq t \leq n_i\} 1 \leq i \leq n\}]$.

**Decryption (CT, SK$_L$):**

The recipient tries to decrypt CT, without knowing the access policy W by using his SK$_L$ associated with the attribute list L as follows

$$M = \frac{C\tilde{~} \prod_{i=1}^{n} e(C_{i,t,2}, D_{i,2})}{e(C_0, D_0) \prod_{i=1}^{n} e(C_{i,t,1}, D_{i,1})}$$

## 4 SECURITY ANALYSIS

**Theorem**: The anonymous CP-ABE construction is secure under the DBDH assumption.

**Proof:**
   We assume that the adversary $\mathcal{A}$ has non-negligible advantage $\varepsilon$ to break the privacy of our scheme.
   Then we can construct an algorithm $\mathcal{B}$ that breaks that DBDH assumption with the probability $\varepsilon$
   Let $(g, g^a, g^b, g^c, Z)$ be a DBDH instance.

**Init.** The adversary $\mathcal{A}$ gives $\mathcal{B}$ the challenge access structure $W_0^*$ and $W_1^*$. $\mathcal{B}$ chooses d randomly from the set $\{0, 1\}$.

**Setup.** To provide a public key PK to $\mathcal{A}$, $\mathcal{B}$ sets $Y = e(g,g)^{ab}$, implies $\alpha = ab$. Choose $a'_{i,j} \in_R Z_p^*$ ($i \in [1,n], j \in [1,n_i]$) and computes $T_{i,j} = g^{a'_{i,j}}$

The simulator, $\mathcal{B}$ sends the public parameters $(e, g, Y, \{\{T_{i,j}\} 1 \leq j \leq n_i\} 1 \leq i \leq n$ to $\mathcal{A}$.

**Phase 1.** $\mathcal{A}$ submits an attribute list L = [L$_1$, L$_2$, …, L$_n$] in a secret key query. We consider only the case where (L ⊭ $W_0^*$ and L ⊭ $W_1^*$).

For KeyGen query L, $\mathcal{B}$ choose $\beta_i, a'_{i,j} \in Z_p^*$ and set $\lambda_i = \beta_i$, $r = ab - \beta_i a'_{i,j}$ and computes the secret keys as follows

$$D_0 = g^{\alpha-r}$$
$$= g^{(ab-(ab-\beta_i a'_{i,j}))}$$
$$= g^{\beta_i a'_{i,j}}$$
$$D_{k,1} = g^{r+\lambda_i a_{i,t}}$$
$$= g^{ab-\beta_i a'_{i,j}+\lambda_i a_{i,t}}$$
$$= g^{\alpha}$$
$$D_{k,2} = g^{\lambda_i} = g^{\beta_i}$$

**Challenge.** $\mathcal{A}$ submits two messages $M_0, M_1 \in G_1$ if $M_0 = M_1$, $\mathcal{B}$ simply aborts and takes a random guess. The simulator flips a fair binary coin d, and returns the encryption of $M_d$. The encryption of $M_d$ can be done as follows:

$$C_0 = g^c \quad, \tilde{C} = M_d\, e(g,g)^{ac} = M_d\, Z.$$

$\mathcal{B}$ generates, for $w_d$, the ciphertext components $\{\{ C_{i,t,1}, C_{i,t,2} \}\; 1 \leq t \leq n_i \}\; 1 \leq i \leq n$ as follows
Set the root node of W to be c, mark all child nodes as un-assigned, and mark the root node assigned.

Recursively, for each un-assigned non leaf node, do the following

a) If the symbol is $\wedge$ and its child nodes are unassigned, we assign a random value $h_i$ $1 \leq h_i \leq p-1$ and to the last child node assign the value

$$h_t = \frac{c}{\sum_{i=1}^{t-1} h_i}.$$ Mark this node assigned.

b) If the symbol is $\vee$, set the values of each node to be c. Mark this node assigned.
Each leaf attribute $w_i$ can take any possible multi values, the value of the share $s_i$ is distributed to those values and compute
$[C_{i,t,1}, C_{i,t,2}] = [\, g^{h_i},\; T_{i,j}^{h_i}\,]$.

**Phase 2.** Same as Phase 1.

**Guess.** From the above considerations, the adversary can decide that $Z = e(g,g)^{abc}$ when $d = d'$ and can decide that $Z \in_R G_1$ otherwise. Therefore $\mathcal{A}$ breaks the DBDH

problem with the probability Є. □

## 5   CONCLUSION

We proposed an Attribute based encryption which preserves the privacy of the access policy, specified by the encryptor. This scheme is very expressive and provably secure under the decisional Bilinear Diffie-Hellman assumption.

## ACKNOWLEDGEMENTS

This work was supported by National Technical Research Organization (NTRO), New Delhi, India.

## REFERENCES

1. Waters,B, (2008) "Ciphertext policy attribute based encryption : An expressive, efficient, and provably secure realization", Cryptology ePrint report 2008/290 .

2. Bethencourt, J,. Sahai, A., Waters,B,( 2007) "Ciphertext policy attribute based encryption", IEEE Symposium on Security and privacy , pp, 321-334.

3. Cheung,, L, Newport,C, (2007) "Provably secure Ciphertext police ABE", CCS 2007: Proceedings of the $14^{th}$ ACM conference on Computer and Communications security, pp.456 -465, ACM Press, New York.

4. Goyal, V., Jain,A., Pandey,O., Sahai,A , (2008)  "Bounded Ciphertext policy attribute based encryption", In: Aceto, L.,Damgard,I., Goldberg, L.A., Halldorsson , M.M., INgolfsdottir,A., Walukiewicz, I .(eds) ICALP  2008, Part II. LNCS , Vol 5126, pp 579- 591, Springer , Heidelberg .

5. Goyal, V., Pandey, O., Sahai,A., Waters,B.(2006) "Attribute –based encryption for fine grained access control of encrypted data",  ACM  Conference on Computer and Communication Security, pp. 89-98.

6. Sahai, A., Waters, B.,(2005) "Fuzzy Identity-based encryption", EUROCRYPT 2005. LNCS,vol 3494,pp. 457-473.Springer, Heidelberg.

7. Nishide, T., Yoneyama, K., Ohta, K.,(2008) "ABE with partially hidden encryptor-specified access structure", ACNS'08, LNCS 5037, pp 111-129, Springer.

8. Emura, K., Miyaji, A., Omote, K.,(2009) "A ciphertext policy Attribute –Based Encryption scheme with strong Recipient Anonymity.",The $4^{th}$ International workshop on Security (IWSEC),49-63.

9. Li, J., Ren, K., Zhu, B., Wan, Z.(2009),"Privacy aware Attribute based Encryption with user Accountability". eprint.iacr.org/2009/284.

10. Boneh, D., Waters, B.(2007) " Conjunctive, Subset, and Range Queries on Encrypted Data". TCC'07.LNCS 4392,pp 535-554. Springer.

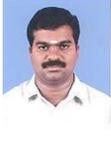 Mr. A.Balu is working as a Research Associate in the project "Smart and Secure Environment", Department of Computer Science and Engineering, Alagappa University, Karaikudi, Tamilnadu, India. He is pursuing Ph.d in the field of Access control Mechanism. He have presented a paper in the International workshop on Trust Management in P2P systems.

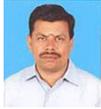 Prof. Dr K.KUPPUSAMY is working as an Associate Professor in the Department of Computer Science and Engineering, Alagappa University, Karaikukdi, Tamilnadu, India. He has received his Ph.D in Computer Science and Engineering from Alagappa University, Karaikudi, Tamilnadu in the year 2007. He has 22 years of teaching experience at PG level in the field of Computer Science. He has published many papers in International Journals and presented in the National and International conferences. His areas of research interests include Information/Network Security, Algorithms, Neural Networks, Fault Tolerant Computing, Software Engineering & Testing and Optimization Techniques.